**Title of the article:**
Commitment Nets in Software Process Improvement

**Authors:**
Pekka Abrahamsson

**Notes:**

# Commitment Nets in Software Process Improvement[1]

Pekka Abrahamsson
VTT Electronics
P.O.Box 1100
FIN-90571 Oulu, Finland
Pekka.Abrahamsson@vtt.fi

**Abstract**

Several studies have revealed the fact that nearly two-thirds of all software process improvement (SPI) efforts have failed or have at least fallen short of expectations. Literature and practice have shown that commitment to SPI at all organizational levels is essential for the success of any SPI endeavor. A research model for studying the existence, development and interplay of SPI-related commitment is introduced in this paper. This study suggests that software organizations operate through strategic, operational and personal commitment nets. These nets consist of actors, drivers, concerns, actions, commitment, and outcomes. The commitment nets model is applied in a study of four industrial SPI initiatives. The results from two of these cases are reported here. The results show that SPI is driven through the formation and reformation of commitment nets. The contents of strategic, operational and personal commitment nets are laid out and implications are discussed.

1. **Introduction**

Software process improvement (SPI) related literature and practice have seldom agreed on the importance of any single concept as consistently as is the case with the commitment factor. Commitment to SPI by all the levels of the organization is one of the most essential factors for determining whether a well-planned process improvement program will succeed or not [Humphrey 1989; Kuvaja, et al. 1994; Dahlberg and Järvinen 1997; Kautz 1999]. Commitment is considered important because it is thought that if an organization and its members are committed to SPI, activities will be sustained even in the face of difficulties, which are often encountered in the case of SPI initiatives. While recent SPI models (e.g. CMM, SPICE, BOOTSTRAP) have paid little attention to human issues [McGuire and Randall 1998], the software engineering community itself has clearly acknowledged the need for gaining a better understanding of the complexities of the turbulent business environment in which process improvement takes place.

A recent analysis on the commitment models that underlie SPI thinking has revealed that these models are based on faulty assumptions [Abrahamsson 2001a, b]. Thus, there is a strong case for creating new enhanced commitment models. These models should be based on empirical evidence and they should not contain any of the pitfalls of the present models. The problem with the existing models, according to Abrahamsson [2001b], is the assumption that commitment would develop neatly in a linear and causal fashion, suggesting that commitment would thus be a controllable, all-positive phenomenon.

This paper seeks to fill this gap by developing a model for studying the existence, development and interplay of the commitment structures involved in SPI. It is suggested that software organizations operate through commitment nets – which can be of strategic, operational and personal nature. The commitment net elements - actors, drivers, concerns, actions, commitment, and outcomes - are introduced. The results show that SPI is driven through formation and reformation of commitment nets. The contents of the personal, operational and strategic commitment nets are then laid out and discussed.

---
[1] An early version of this paper was presented at the Hawai'i International Conference on System Sciences (HICSS35), 2002.

This paper is organized as follows. First, an overview on the conceptual and processual aspects of commitment related phenomena is provided. Based on this, a research model for studying commitment in software process improvement is introduced. This is followed by a description of the research design. Then, two software process improvement cases are analyzed using the commitment nets model and the results are presented. Finally, a discussion of the implications of these results is provided. The paper concludes with a section for final remarks.

## 2. Background

Commitment has been one of the most popular research subjects in industrial psychology and organizational behavior over the past 30 years [Benkhoff 1997]. The reason for a widespread interest on the subject has been the assumed relationship between commitment and performance. However, no such direct relationship has been found [Mathieu and Zajac 1990]. In fact, only a few alleged impacts - i.e., consequences - of commitment have been empirically validated even though hundreds of studies have made an attempt to examine the correlations between commitment and different variables [Meyer and Allen 1997]. The best results in studies concerned with commitment to one's organization are achieved with predicting turnover rates [Mathieu and Zajac 1990]. It is, however, not easy to provide a clear-cut definition of commitment due to the fact that there are a number of different interpretations of the concept. The reason for this inconsistency and confusion has been attributed to the lack of a specific commitment model [Coopey and Hartley 1991]. In the following, a brief conceptual synthesis and a look at the process aspects are provided.

### 2.1. The Concept Commitment

Commitment is a state of attachment that defines the relationship between an actor and an entity [O'Reilly and Chatman 1986]. The actor may be a single individual, a group of persons (e.g. project team) or an organization [Newman and Sabherwal 1996]. The relationship can be viewed in terms of strength, focus, terms [Brown 1996], durability [Abrahamsson 2001b] and component type [Meyer and Allen 1991]. These aspects are common to all commitments. Table 1 briefly describes the different aspects of commitment.

Table 1. Common aspects of commitment.

| Aspect | Description | Expected in SPI |
|---|---|---|
| Form | The nature of commitment: Commitment is formed of its components. At least four forms of commitment exist: Affective, normative, continuance and instrumental. These forms build up a composite that changes over time. These varieties may also be seen as sources of commitment, i.e. motives engendering attachment. [O'Reilly and Chatman 1986; Meyer and Allen 1991; Becker 1992] | Existence of affective, continuance, normative and/or instrumental forms of commitment toward SPI. |
| Focus | Defines the target of one's commitment. The target can be work or non-work related. Work related targets are, e.g. organization, project, one's career or profession. [Simon, et al. 1950; Gouldner 1960; Coopey and Hartley 1991; Morrow 1993; Brown 1996; Meyer and Allen 1997; Baruch 1998] | Commitment to SPI appears during the SPI project. |
| Durability | Depending on the commitment target, the durability of 'personal contract' varies. For example, commitment to career may last for a lifetime but commitment to a project does not. [Abrahamsson 2001b] | Commitment to SPI is dependent on project's duration. |
| Terms | Terms of commitment define what has to be done in order to fulfill the requirements of a commitment. A contract, for example, is an explicit pact where the terms are stated. [Brown 1996] | Commitment to SPI is reflected through acceptance of goals, willingness to exert extra effort, and persistence in achieving the set goal. |

| Aspect | Description | Expected in SPI |
|---|---|---|
| Strength | Defines how deeply a person or a group is attached toward an entity. A person is likely to be more or less committed, rather than simply committed or not. [Kiesler 1971; Brown 1996; Beck and Wilson 2000] | Different levels of commitment to SPI exist. |
| Actor | Unit of analysis in commitment studies. 1) An individual, 2) a group or team or 3) an organization can show commitment toward an entity. [Newman and Sabherwal 1996] | Depending the on the actor, all the common aspects introduced above may vary. |

While most aspects of commitment are essentially self-explanatory, the component form requires further explanation. Although components rarely exist in their pure form, they are referred to as archetypes of commitment [Abrahamsson 2001b]. Depending on the target of commitment and circumstances, the SPI-related commitment of a software developer may manifest itself in all the forms described above. One of the reasons may be that since they *want to* take part in the initiative, they see its benefits (affective component). Secondly, they may feel they *need to* continue participating, because they have invested a lot of their time in the endeavor, and they may also need to see the results in order to justify the rationality of SPI activities (continuance component). Finally, the organization, the project manager and the SPI staff expect results from the initiative, which creates a sense of obligation either to participate or to demonstrate results (normative component).

Participation in an SPI activity in itself is not necessarily enough to demonstrate that a person is committed to SPI. Committed behavior expends effort "beyond contract" for the enterprise [Bratton and Gold 1999]. Commitment researchers link commitment with an idea of a worker willing to "go the extra mile" on behalf on the organization [Mowday, et al. 1982]. This is especially the case with the affective form of commitment. Accordingly, commitment researchers have suggested that affective commitment is the most desirable form of commitment [Meyer and Allen 1997]. Abrahamsson [2001b] has argued similarly for SPI-related commitment.

## 2.2. Commitment process

While the concept of commitment can still be considered slightly ambiguous, even less is known about the process itself. Commitment researchers' self-critique has revealed that they have not been paying too much attention to the process itself [Staw and Salancik 1977; Scholl 1981; O'Reilly and Chatman 1986; Meyer and Allen 1991]. The few considerations that do exist are, for the most part, of speculative nature.

> "Considering the paucity of studies, however, this discussion is necessarily speculative. It is intended primarily to illustrate the importance of process considerations, to indicate how different processes are likely to operate for affective, continuance, and normative commitment, and to provide some direction for future research." [Meyer and Allen 1991, p.74]

While it is not clearly known how commitment develops, we have an idea how it doesn't develop. Thus, if these faulty assumptions in current thinking are explicated, we will be in a better position to avoid them. In what follows is a brief synopsis of the main findings made in a recent study [Abrahamsson 2001b] concerning the critical misconceptions underlying current SPI thinking (Table 2).

Table 2. Assumptions underlying SPI research and practice

| Assumption | Current practice | Example |
|---|---|---|
| Causality in human cognitive processes | Commitment models are linear with a definite set of steps that everyone goes through | [Conner and Patterson 1982] |

| Assumption | Current practice | Example |
|---|---|---|
| The controllability of this process | Tactics or cookbook type recipes on how to get someone committed or how to handle the commitment problem. | [Drennan 1989] |
| The notion of singular commitment construct | There is only one type of commitment and this commitment is 'needed' in SPI | [Humphrey 1989] |
| The idea that commitment is an all-positive phenomenon | Commitment-oriented culture is needed for implementing SPI | [Hadden 1999] |

Thus, as is indicated in Table 2, these underlying assumptions are likely to direct SPI related research and practice. They are lead by a lack of studies and they have resulted in an oversimplified - yet attracting - view on commitment. This attraction is due to the allegedly influential role of commitment in SPI. While the concept remains unclear, it also starts to take on mysterious meanings such as the notion that commitment is way of life [Humphrey 1989] or what actually is needed is a commitment-oriented culture [Hadden 1999]. A recent SPI implementation model defined commitment as one separate element of its own [Isacsson, et al. 2001].

How does, essentially, commitment develop then? A lot of the research carried out on commitment has been variance theory oriented, suggesting that as soon as certain conditions are met, commitment [to any target] will appear [Montealegre and Keil 2000]. This has been the case, even though it has also been proposed that commitment is a continuous variable rather than a dichotomous one [Kiesler 1971], i.e., people are referred to as being more or less committed rather than being simply committed or not. While the research should move beyond the state of "a simple laundry lists of antecedents and correlates", as Meyer and Allen [1991] have put it, a new model for studying the existence and the development of commitment is needed. To serve this need, this paper introduces the model of commitment nets. The proposed model is introduced in the following section.

## 3. Model of commitment nets

Inspired by the writings of Winograd, Flores and Spinosa [Winograd and Flores 1987; Flores and Spinosa 1998] the idea of *commitment nets*[2] emerged. Flores draws on the works by Heidegger (1962 [1937]), Kierkegaard [1985] and Hegel [1979]. Rather than giving any detailed description of the philosophical underpinnings of their writings, the intent here is to further elaborate the interpretation of Flores et al. in relation to understanding commitment and its operation in organizations. Winograd and Flores [1987] use the ideas of speech act theorists [Austin 1962; Searle 1979] to demonstrate that "human beings do not normally act in the world by simply transferring, disassembling, and reassembling basic thing". Flores calls this "The Cartesian misunderstanding of language and communication". Winograd and Flores argue that this misunderstanding has damaged our productive capacities, compromised our ability to recognize ourselves inside a changing community, and caused a failure to cultivate our political capacities.

> "Because we tend to look at components of things and not commitments and how they are structured with each other, we increasingly find that our domestic and commercial lives are transforming themselves beyond our control." [Flores and Spinosa 1998, p.355]

---

[2] The concept "net" refers to the relationships between different actors (1st order net) as well as the interrelationship of the elements (2nd order net) in the [commitment] net itself.

Winograd and Flores have "opened the discipline of tracking, mapping, and combining commitments based on the constituting power of human speech". In their writings on computers and cognition [Winograd and Flores 1987], they have stated that there is a general structure for forming commitments for actions to satisfy concerns. Concern, to them, is an ongoing generalization of a need. Through the focus on concerns and commitments, new domains of assessment emerge. One of these domains is "the identification of the new institutions that are arising alongside old ones". This is explained as follows:

> "Mapping social institutions in terms of their concern and commitment structure tells us what is genuinely new and what is a new way to accomplish old goals" [Flores and Spinosa 1998, p. 357]

In our approach, the term social institution refers to the way software is produced in a software-intensive organization. A project can be considered another form of social institution, with its specific concerns and commitments. If this is accepted, SPI can be seen as a method for changing or altering this institution. If we are to evaluate whether a change has occurred, we have to look at the changes in concerns and commitments, i.e., the changes taking place in the respective actors' commitment nets.

> "[…] once we become familiar with the way commitments drive action, we no longer believe that we have to understand in advance all the component parts of whatever social action we seek. Rather, we see that we must identify concerns and begin forming commitments to address them. The basic organizing skill is forming and managing a commitment to deal with a concern. On the basis of one commitment, many others can be grown" [Flores and Spinosa 1998, p. 357]

The key concepts constituting the basis of the *commitment net* (Figure 1) are actors, drivers, concerns, actions, commitments and outcomes. These concepts are introduced in the following sections. Majority of the discussion presented here focuses on the individual level, referred to as the *personal* commitment net. Inter-organizational clusters and the commitment nets of individual organizations, for their part, are strategic in their intent, which is why these levels are referred to as *strategic* commitment nets. An organization operates through different groups, which are referred to as *operational* commitment nets. The common aspects of commitment introduced in the earlier section (form, focus, durability, terms, strength) provide a typology for studying the existence of concerns and actions in a commitment net.

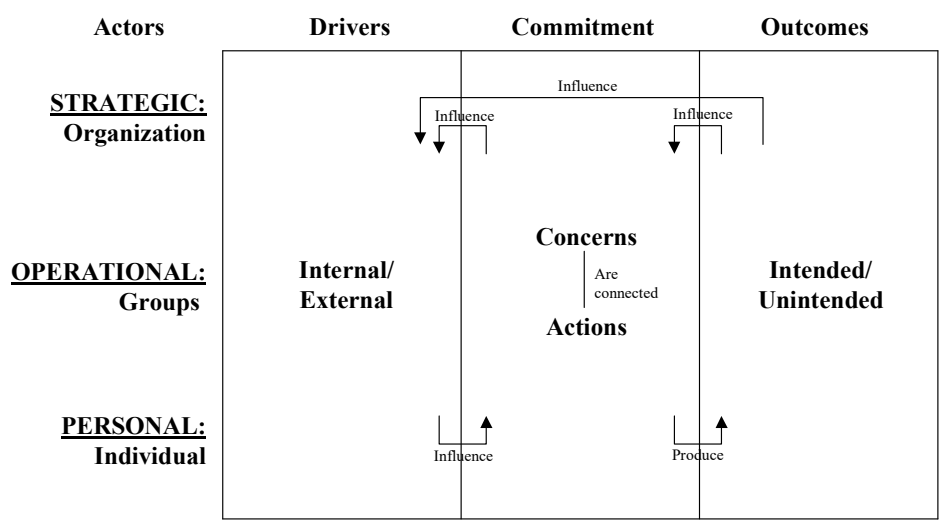

Figure 1. Commitment nets model

### 3.1. Actors

SPI efforts are rarely, if ever, designed for a single[3] software engineer, but most often for a group of engineers or for an entire software development organization. Acknowledging the fact that organizations operate in a networked fashion [e.g., Seppänen 2000], in which the actors and their relationships constitute the elements of the network, these institutions can be considered to operate at different levels. There are three levels of actors: individual, group and organization. These actors form three levels of commitment nets, respectively: strategic, operational and personal. The actors in SPI context are mapped with actor groups in Table 3.

Table 3. Actors in SPI context.

| Actor | Role in SPI | Function in SPI |
|---|---|---|
| Organization (strategic) | Sponsorship | Authorization of budgets and resources |
| | Management | Provides management guidance and strategies; monitors progress; resolves organizational issues; promotes SPI goals |
| Group (operational) | SPI team | Manages and implements process improvement activities |
| | Object of change: a Software project | Participates in the change; adopts new behavior patterns or tools |
| Individual (personal) | Key stakeholders | Key people from either of the actor groups. Facilitates SPI progress. |

### 3.2. Drivers

Research in related disciplines [cf. Sabherwal and Elam 1995; Meyer and Allen 1997] has shown that there are a number of drivers contributing to the creation of concerns and actions (i.e., determinants of commitment). These drivers are either internal or external to an actor [Nijhof, et al. 1998]. An exhaustive literature review has revealed over 70 possible determinants for commitment development. Scope and target of this paper make it unnecessary to describe each determinant in detail. Instead, only the drivers identified in IS commitment research [Sabherwal and Elam 1995] that have been adopted for the initial model are introduced. Similarly to information systems development, SPI activities are most often performed as projects with their own budget, timescale, dedicated resources and goals [Zahran 1998]. Four driver categories internal to an actor can be identified: psychological, project, social and structural drivers. These categories are briefly described in the table below.

Table 4. Drivers for commitment

| Driver | Description |
|---|---|
| Psychological | These drivers involve key individuals in the project reflecting properties such as need for achievement, past historical success, etc. |
| Project | Reflects objective features of SPI project in terms of costs and benefits. Project drivers indicate reasons for an SPI to exist |
| Social | Social drivers originate from a group rather than an individual. These drivers are, for example, power and politics, or public identification with the SPI project. |

---

[3] A notable exception to this is the Personal Software Process (PSP) method [Humphrey 1995], which is explicitly targeted to a single software engineer.

| Driver | Description |
|---|---|
| Structural | Structural drivers represent the contextual conditions surrounding the project: the environment for SPI activities. |

External drivers are, e.g., company culture, software engineering profession, social trends, marketplace and the economy [Sharp, et al. 1998]. It is difficult for an actor to directly influence external drivers. However, external drivers may have a strong influence on actors' concern and action creation.

3.3. Commitment – concern and action

The concept of commitment as used in this paper differs from that of Winograd, Flores and Spinosa [Winograd and Flores 1987; Flores and Spinosa 1998]. While they suggest that commitment can be separated from concern (by stating that commitments are formed to satisfy concerns), we maintain that those two are inseparable if we understand commitment as a state of attachment. A concern without corresponding action cannot be regarded as commitment. Consider the following example.

> A person argues that he is committed to preserving the nature. One would then assume that this person acts in a manner that would indicate the existence of such a commitment. He would recycle waste, attend environmental group meetings or give out pamphlets in the street. However, if he does not perform any actions matching his concern with nature preservation, one would conclude that such a commitment does not truly exist.

Action, in itself, does not bring about commitment, either. However, action should be interpreted as a sign of potential commitment. If the person in the example above would indeed stand in the street on a rainy day distributing pamphlets about nature preservation, one would conclude that this person is committed to the cause. The cause, however, is not clear to the observer. The observer's *perception* of the situation (rainy day, contents of the pamphlet, etc.) and action (act of distribution) contributes to his understanding of a possible commitment target. One may, indeed, interpret that the cause is nature preservation. Still, the observer does not know why this person is committed to the specific cause. Here, we are able to benefit from the archetypes (or forms) of commitment. We maintain that the person is committed to the cause in all the forms of commitment as proposed previously. In many cases, however, understanding the nature of a commitment is not as important as the fact that there is such a commitment present. Moreover, the cause of a perceived commitment is not necessarily accurate, since it might be, for instance, that the person in our example is just performing the act of distributing pamphlets due to fact that this person needs some extra money to support his family. Thus, the question arises what the concern driving the action is, which, in turn, will reveal the true commitment target to the observer. The behavioral school (see e.g. [Mowday, et al. 1982] for details on the different perspectives) of commitment research emphasizes the role of action as commitment target.

> "[…] commitment targets should be actions rather than objects, as it is virtually impossible to describe commitment in any terms other than one's inclination to act in a given way towards a particular commitment target" [Oliver 1990, p. 30]

To some degree, the argumentation of the behavioral school appears acceptable. However, we are proposing that action, in itself, can be regarded as no more than a potential commitment target. If it were to be considered commitment, it should also hold the state of attachment (i.e., concern). A recent study by Shephard and Mathew [2000] explicates that 97,9% per cent of respondents agree that attitude differentiates committed employees from non-committed ones, general behavior was stated by 72,3% of the respondents as an indication of commitment. Thus, it is only by discovering concerns and related actions in a software organization at the different operational levels that enables us to identify potential commitment targets. This knowledge puts us in a better position to analyze the arrival of new commitment – i.e. commitment to SPI in this study. A lot of anecdotal evidence shows that such commitment does not necessarily come about.

3.4. Outcome

Only action produces outcome. Orlikowski [1992; 1993] in her famous study on implementing the Lotus Notes groupware tool in an organization distinguished (among others) between intended and unintended outcome, which was produced by this groupware software for the organization. Similarly, SPI activities lead to intended and unintended outcome, which both in turn affect the drivers, concerns and future actions. In SPI, unintended outcome may take the form of dissatisfaction with an introduced tool, which may subsequently lead to difficulties in co-operation between process and project departments. Unintended outcome may also be positive. For example, improved work morale or a range of new SPI activities (that were initially not intended) may emerge from an SPI effort. Intended outcome reflects the fact that the SPI project has achieved the goals set for it. In the empirical part of the paper, the outcome is regarded as issues, actions, artifacts, etc. produced by the SPI project being studied. While action produces outcome, the outcomes has an influence on its operating environment – i.e., on commitment and drivers.

**4. Research design**

4.1. Research methodology

Action research was employed in this study as research methodology. Action research produces knowledge for guiding practice [Oquist 1978], which was the aim of this study. The purpose of the researchers was to gain enough knowledge of the existing situation, after which changes to reality (or current practice) were proposed. In Lau's [1997] contemporary IS Action Research Framework, the type and focus of research, along with the underlying assumptions and the research process must be documented. Here, special attention is paid to the type and focus as well as to the research process itself.

Using an action research cycle as proposed by Susman and Evered [1978], the change effort was guided through. A single cycle involves five stages: 1) diagnosing the problem, 2) action planning, 3) action taking, 4) evaluating and 5) specifying learning. However, since a single cycle may take from months to one year to complete, as Kock [1997] has recently proven, a second cycle was implemented only in one out of four cases. In addition, since a single change effort can rarely be isolated from the larger picture, no rigorous separation was enforced between the different stages. Also, the level of involvement differed from mere observation to actual implementation. Using the taxonomy proposed by Baskerville and Wood-Harper [1998], the main type of action research employed was that of participatory action research, in which the investigator participates (or intervenes) in organizational daily work and treats subjects as equal co-workers. The focus is on organizational development and advancing scientific knowledge on the subject matter for academia. In our case, advancing scientific knowledge implies the effort to explore and to understand the process of commitment in the context of software process improvement effort. While a series of efforts (use of diary, for example) were made to involve practitioners as co-researchers, this was just marginally achieved only in one case. Practitioners are often too busy to stop to reflect upon their work.

4.2. Research setting

Two organizations were involved in this study. The first organization (Company A) is an SME (Small to Medium Sized Enterprise) developing electronic ticketing systems for national and international markets. Company A is divided into five different business units: after sales, projects delivery, product development, systems unit and marketing. The second organization (Company B) is an independent business unit of a large global corporation that develops electronic products and systems for industrial customers. It has thousands of employees working in several locations around the world. Company B has been organized into a matrix, where the line organization is responsible for resource management and competence development, and the project organization takes care of product development.

Four cases were selected for this study. Two of them, cases 1 and 2, are focused on in this paper. Figure 2 demonstrates how the cases map to the software engineering paradigm. The V-type software process development model[4] serves here only as an example. It is not maintained that the organizations involved in the study did actually employ such a process model in their software development. Figure 2 is thus divided into four process sections: user-centered/service process, software development process, project management process and SPI process. The user-centered/service process also includes customer support activities. The software development process describes activities taking place during software development, including contracting, requirement analysis, system design, program design, coding and unit testing, integration testing, system testing, acceptance testing, along with operation and maintenance. The project management process covers management activities related to software development, including project planning, scheduling and risk management. Software quality assurance processes typically involve activities aimed at ensuring the quality of project outcome. SPI is a support process typically involving quality assurance activities along with giving concrete support for software development projects. [Sommerville 2001]

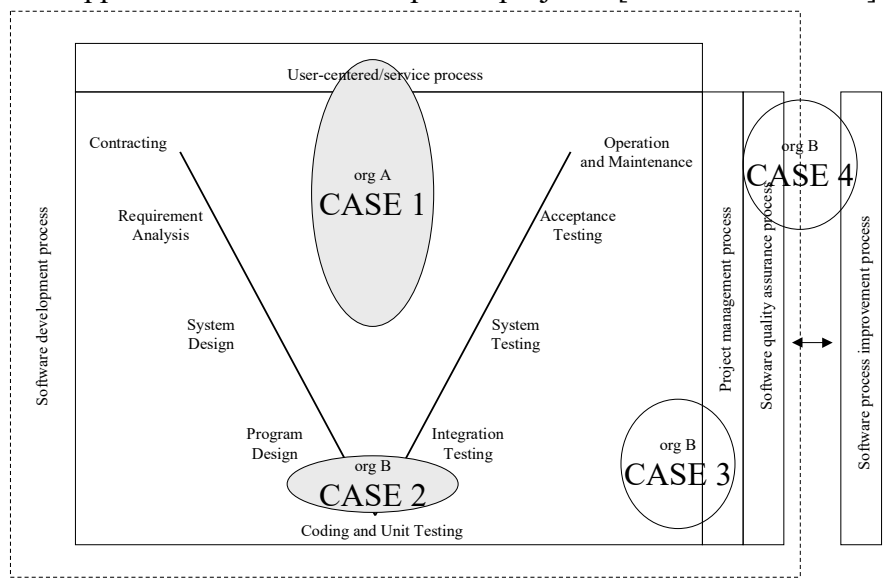

Figure 2. The position of the cases within the framework of software development.

To complement the picture given above on the relations of the cases treated in this study to the software development process, the following table provides some basic information about the cases: goal, starting and ending time, scope, type of SPI effort and researcher's role. The type of SPI effort is adopted from Seppänen et al. [2001]. According to the authors, there are four possible SPI types: process repair, optimization, re-engineering and innovation. Process repair refers to seeking short-term gains through a special solution. For example, problems with the existing process are solved by acquiring a new tool. Process re-engineering, in turn, usually focuses on actors, revising their organization and responsibilities in the processes. Process optimization, for its part, lays emphasis on activities. According to our interpretation of this type, the current process already produces acceptable results, while enhancements are made to increase its effectiveness or speed, or both. Finally, process innovation focuses on the interplay of several processes rather than on some actors, activities or resources. Seppänen et al. [2001] argue that process innovation is likely to bring the most value to an organization.

---

[4] The V-model is a software development standard that addresses SE lifecycle and its supporting activities. It is a national standard for use of the German Federal Armed Forces. The original description of the model can be found at [IABG 1992] and a summary in a recent work by Zahran [1998, pp. 377-386].

Table 5. Summary of the cases

| Case id | Goal(s) | Duration | Scope | Type | Researcher role |
|---|---|---|---|---|---|
| 1 | Spread usability thinking; adopt UCD practices | 14 months (ongoing) | Organization-wide | Process innovation | Expert consultant |
| 2 | Improve module testing practices | 12 months | Individual software designers | Process repair & re-engineering | Expert consultant, participant observer |

4.3. Data Collection and analysis

A wide variety of research material was collected: slightly over 1000 pages of transcribed text from over 30 open-ended interviews, company internal documents, research diaries, notes from corridor discussions and emails, along with a survey on module testing development in case two. The material was organized on a case-by-case basis and each case was analyzed separately using the commitment net model presented earlier in this paper.

Since process data is often difficult to interpret, various research strategies proposed by Langley [1999] were employed. More specifically, the narrative approach was chosen for building a narrative description of events that took place in each initiative. This was used for communicating the results to the participating organization, thus also ensuring the use of the member-checking method. The grounded theory strategy was used for distinguishing the elements of commitment net – drivers, concerns, actions, and outcomes.

## 5. Empirical case analyses

This section comprises the empirical part of the paper. Two software process improvement cases are analyzed using the commitment nets model introduced above. Both cases are described in three sections: project initiation, operation and closure. In the initiation phase, the actors involved are identified along with their driving interests. The initiation phase also involves assessment activities (case one). The project operation phase comprises implementation activities. Finally, in the closure phase, the analysis evaluates the extent of change achieved in the commitment nets of the various actors involved.

5.1. Case 1: Improving usability processes

Company A participated in a university driven research project. The aim of the project was to enable companies to enhance their competitiveness by producing scientifically and empirically validated models for business-based user-centered software development. This study evaluates the progress from February 2000 to April 2001. The improvement approach was changed at that time.

### 5.1.1 Project initiation

Company A was reorganized a year before the usability initiative was launched. Due to this, the company atmosphere had deteriorated considerably and the management team feared that this would reflect their image outside. In addition, an ongoing large project had been postponed several times due to problems with defining a number of user-interface issues in the product. The management thus saw many opportunities in participating in the usability improvement project. In their opinion, stressing the usability issues would benefit the company in a number of ways: in addition to improving the company image (recruitment perspective), usability would give them a competitive edge over competitors (business perspective) and it would give them control over the customer in defining the user interface requirements (development perspective). These drivers are all of structural nature and, accordingly, they had a positive effect in the initiation phase. Furthermore, the participation in the project was seen as cost-effective – which can be considered a project driver – since the services and guidance in the project they were to receive from the university would by far exceed the participation costs.

Company A has had a long history of high success in producing products that had been complemented on their ease-of-use. Thus, there was also a psychological driver affecting the strategic commitment net positively. While the participation was seen as a cost-effective action, the intent of the usability project itself was not, however, well understood.

> "[…] [Usability improvement project] was seen as a part of a university project. This meant that it was too theoretical […] We launched a project but did not think what benefits would be gained. We let the university lead it." [Management team member, interview]

As a part of the initiation phase, a two-week long usability capability assessment was conducted, including the introductory and closing sessions. In the early days of the project, this received high priority among the actions of the usability improvement project. However, the assessment procedure proved to cause more confusion than clarity concerning the position and status of usability in the organization.

> "It contained a lot of unfamiliar terms both to me and to the rest of the audience. So we did not understand too much of the presentation. […] I told them[personnel] that of course there are a specific terminology and other stuff related to usability. There's some kind of mismatch between the academia and the software development organization. Here we aim at developing concrete products, this is not a research organization. […] Some of the personnel asked me if I did understand what the researchers were talking about. They used such weird language, how could you cooperate with them. [Usability specialist, interview]

As a result, the assessment team constructed a report of the current status, but failed to come up with any improvement plan. The usability capability assessment resulted in alienating the staff from usability thinking. Thus, it had a negative impact on the project. A project driver also had a negative impact on the commitment at the strategic level. Although operational level actors – systems, mechanics and after sales units – did have participants in the internal usability project of the company, they were treated as project resources rather than representatives of their units, which was the original purpose. To them, the goals and the current state diagnosis actions of the usability project remained unclear. Thus, the project driver had a negative impact on their commitment development.

While all the project participants have their personal commitment nets, this analysis considers only the key player's personal commitment net, i.e. that of the usability specialist within the organization. She showed a clear interest toward usability issues and she had also participated in a number of university courses dealing with various usability aspects. In her studies she had reached the point where she could start writing her master's thesis and she was offered a position as the project manager for the usability project, which she saw as a career development opportunity. Thus, all these elements represent psychological drivers affecting her commitment development positively. Table 6 summarizes the drivers at the project initiation phase.

Table 6. Case 1: Commitment net drivers in the project initiation phase.

| Commitment net level | Actor | Driver | Type |
|---|---|---|---|
| Strategic | Organization: Management | Recent changes in the organization<br>Deteriorated company image<br>Problems with customers (concerning user requirements)<br>Need for competitive advantage | Structural (+) |
| | | Cost-effective participation in the project | Project (+) |
| | | Project goals unclear, assessment results incomprehensible | Project (-) |
| | | Long history of usable products | Psychological (+) |
| Operational | Group: Systems unit, mechanics unit, after sales unit | Project goals unclear, assessment results incomprehensible | Project (-) |
| Personal | Individual: Usability specialist | Interest in usability issues<br>Participation in university courses<br>Preparation of Master's thesis<br>Availability of project manager position | Psychological. (+) |

### 5.1.2 Project operation

An improvement initiative was thus launched to incorporate state-of-the-art usability practices in the case company A. The usability project team was established to implement and to control improvements. A usability specialist was appointed as the project manager. Several actors were involved in the initiative: the management team was an organization level actor, while software projects, mechanics, after sales and the research team from the university were operational level actors. The usability specialist was the key actor at the individual level. In Table 7, the progress of the usability improvement project is presented in a chronological order using the commitment net typology.

Table 7. Case 1: Commitments reflecting the progress of the improvement project.

| # | Actor | Commitment | | Outcomes | |
|---|---|---|---|---|---|
| | | Concern | Actions | Intended | Unintended |
| 1 | Organization: management | Quality (usability) problems in the products, company image, usability thinking missing | Participation in the inter-organizational usability improvement project, establishment of company's internal usability project | Internal usability improvement project established, goals and mission defined | Company was defining actions on their own without consulting with the university research team |
| 2 | Group: Usability project team, researchers | Current status of UCD in the organization, areas for improvement | Usability capability assessment performed, emphasis on the role of management in the assessment | 20 persons interviewed, a report of the current status produced, management had a visible role in the assessment, all senior management participated in the assessment | UCD was perceived difficult to understand and too theoretical, an improvement plan was not produced, staff alienated from usability thinking |
| 3 | Group: Usability project team | Usability related standards should be used for convincing the customer, need of in-house style guidance | Review of usability related standards | Preliminary reports produced, company refers to the use of standards in their web pages | It turned out to be difficult to take bits and pieces out of different standards and put them in operational use. |
| 4 | Group: Usability project team | Usability requirements for the new product, master's thesis | Customer visits with the usability project team | Customer data gathered about current problems and work procedures, but no report produced | Caused problems with personal relationships (some members did not deliver their observations.) |
| 5 | Group: Researchers | Produce concrete and meaningful benefits for the company, ensuring financial support for company A in the future | A paper prototyping course organized | Efficient teamwork, specification of user interface for the new product feature produced, paper prototyping continues to be used in the company | The course proved to be the main reason for staying with the university driven inter-organizational usability project |
| 6 | Organization: Management | Usability project exists only for the preparation of a single master's thesis, no return on money and resources spent, usability project is operating in isolation | A decision to continue financing the usability improvement project: Questioning the benefits gained from the project in management and usability project steering groups | An analysis report of benefits gained produced -> A decision to continue participation within the inter-organizational usability project | The emphasis on usability gained 'too much' attention; expectations raised high. |
| 7 | Group: Systems unit | Usefulness of paper prototyping for requirements specification | Questioning the reliability and extensiveness of the specifications produced through paper prototyping | Specification carefully reviewed regarding concerns made by the systems unit manager | Hostile attitude developed between the systems unit (SW projects) and the usability project |
| 8 | Group: After sales unit | After sales unit had not received any benefit from the usability project | Enhance involvement in the project through joint meetings and workshops | Usability requirements for the user manual and for the implementation process of the new system defined | After Sales unit manager starts to consistently emphasize the importance of the usability project |
| 9 | Focal organization: Management | Usability of the new product | Concentrate all the effort in the usability improvement project on new product development, demonstrate new product in the bus marketing congress | New product designed, highlighting the usability requirements and user feedback, positive customer feedback | The term "usability" gains negative connotation in the organization: "It blocks the new product development schedule" |
| 10 | Group: Researchers | Inefficient collaboration with systems unit | Implement an evolutionary improvement approach | Not included in the analysis | Not included in the analysis |

### 5.1.3 Project closure: Changes achieved

To make the usability improvement project succeed, commitment nets at all levels should be changed permanently to include a concern about usability, along with actions to match this concern. While several SPI actions were employed to incorporate UCD into the daily routines of company A, after fourteen months into the improvement project UCD practices are still not part of the daily routines. Table 8 summarizes the extent of changes, i.e. the SPI outcome, achieved in the actors' commitment nets.

The usability specialist's devotion to the project provided the greatest contribution to the progress of the improvement project. When asked whether usability activities would be sustained if she were to leave the organization, she replied:

> "But [usability activities] would not be continued at this level. Some actions would be carried out, but the whole concept of user-centered design would stay in the dark." [Usability specialist, interview]

She was working as the project manager of the usability project and the leader of the user interface team. She was able to deliver 100% of her effort to these roles and has shown strong personal interest in usability issues. She, for example, took the time in her summer vacation to visit a national bus exhibition just out of pure interest. Her commitment toward the usability project was so intense that she even performed tasks that others were supposed to do.

Another commitment net that contributed to usability project during the period of investigation and was also successfully altered was the strategic one. The founder (and co-owner) of Company A participated in the project in a consistent and active manner. He took an especially active role in usability assessment, even though he did not thoroughly understand what the assessment was all about. He continued to publicly emphasize the importance and benefits of the usability project. The interest of the management in the usability project was also based on the need of an enhanced competitive edge. The Internet pages of Company A now indicate that they adhere to the ISO 13407 (Human-oriented design processes for interactive systems) and ISO 9241-11 (Guidance on usability) standards. Thus, the change in strategic commitment net is clearly visible to other organizations as well as to Company A personnel.

While the usability improvement project has been able to contribute to the strategic and personal commitment nets, the operational level, however, has proven to be the most difficult to change. Originally, we, i.e. the research team, had thought that the most important commitment net to influence was the strategic one. This was due to the need of securing a continuing financial support and participation for the usability project. However, improvements that are intended to be sustained even if some individuals should relocate themselves have to be implemented at the operational level – more specifically, in the systems unit in this case. These realizations lead us to propose a change, from norm-based to evolutionary, in the approach of the usability improvement project. As stated earlier, however, this falls beyond the scope of this analysis.

Table 8. Case 1: Changes achieved in actors' commitment nets.

| Commitment Net | Target actor | SPI outcome | Evidence |
|---|---|---|---|
| Strategic | Organization: Management | Usability seen as one tool in management toolbox. Used predominantly as a marketing device. | Usability used as a marketing tool, web pages state the use of usability standards, continuing participation in the inter-organizational usability project. |
| Operational | Group: Systems unit | Systems unit alienated from the usability improvement project; apart from the paper prototyping method having been used occasionally, UCD practices have not been diffused. | Unwillingness to participate in joint workshops, group members claim that usability and UCD are only theoretical matters, thus not related to their practical development work. |

|  | Group: After sales | After sales unit manager convinced of benefits, continues to emphasize the importance of the project publicly. | Joint workshops organized; usability requirements for user manual and its implementation process defined. |
|---|---|---|---|
| Personal | Individual: Usability specialist | Training, skills and experience gained in relation to UCD. Attitude enthusiastic toward the usability concept. | Active and enthusiastic participation in the project. |

## 5.2. Case 2: Improving module testing processes

Company B has a long history of producing software on their in-house platform. A high management decision was made stating that the new product would be implemented on a commercial platform. The new platform required the adoption of new programming methods, tools and processes. Thus, also low-level practices were to be changed. This also involved redefining and distributing module testing (MT) practices. This analysis concerns the period of 12 months when several strategies were employed to achieve the set target.

### 5.2.1 Project initiation

The process department had been discussing the defining, improving and harmonizing of module testing practices with project teams for some time already. In fact, the SPI project manager had reported that he had been defining and redefining module testing practices for three years at the time of the interview. For historical reasons, each development unit had their own ways of performing the module testing procedure. This practice showed differences also at the individual level, where different designers had been responsible for their own modules. Company B was in a transition phase where new product development upon a new operating platform was well under way. The intention was to reduce the software development cycle time within the new development paradigm from twelve to nine months. The module testers in the software team had been complaining about the inefficiency of current techniques for a long time already. Each team had solved the problems face as they had seen fit. The process department was not aware of the tools that were currently used. Furthermore, the environment that they were to support had changed in a few years from a single product to multiple products and operating systems. At that time, a total of four testing environments had to be supported. Thus, the problems concerning module testing improvement were widely recognized. No new practices for the new technology or for the software paradigm had not yet been defined. Little change resistance was therefore expected.

> […] Well, now there is an opportunity to do again. New product [development] is underway. This means that [improving the module testing methods] should be done now or it will never get done. Now we have the time to do it. Now we can fix it for good, as otherwise there will never be enough time, and for sure, no second chance will be given after [some solution] has been implemented. [Software designer, interview]

Table 9. Case 2: Commitment net drivers in the initiation phase.

| Commitment net level | Actor | Driver | Type |
|---|---|---|---|
| Strategic | Organization: Management | New product development, testing took over half of the total development time, software development cycle should be reduced by 30% -> Testing process will be automated to the highest possible extent (also MT phase) | Structural (+) |
| Operational | Group: Software projects | New product development, not enough support available<br>Module testing difficult and time consuming, timing is right | Social (+)<br>Psychological (+) |

| Commitment net level | Actor | Driver | Type |
|---|---|---|---|
| | Group: Process department | Poor history of success in the development of module testing practices | Psychological (+) |
| | | Too many tools & platforms to support, new product development ->Increasing need for complete support also in MT methods, tools and processes | Structural (+) |
| | | Unawareness of current status in terms of module testing practices | Project (+) |

### 5.2.2 Project operation

No single overarching initiative was launched to tackle the module testing problems. Instead, different strategies – i.e. different levels of actions – were employed to define and to introduce new module testing practices to Company B. The following strategies were thus employed: (1) Develop new process descriptions and operating instructions, (2) define new roles and responsibilities, (3) evaluate a module testing tool with volunteer SW designers, (4) develop and distribute a MT survey to software designers, (5) evaluate module testing tools and practices in a pilot project, and (6) hire a researcher to evaluate reasons for the little progress made in improving the practices. The progress of the SPI effort is laid out similarly as in the first case in Table 10.

Table 10. Case 2: Commitments reflecting the progress of improvement project.

| # | Actor | Commitment | | Outcomes | |
|---|---|---|---|---|---|
| | | Concern | Actions | Intended | Unintended |
| 1 | Group: Software project, process department | MT instructions outdated | Develop common MT process and instructions | Common MT process defined, MT instructions revised, MT reporting practice redefined | Input provided for a larger test process improvement initiative launched in the new organization. |
| 2 | Group: Process department | Inability to provide adequate support | Define new roles and responsibilities | MT Power User allocated in each software development team, MT Core Team established | Change perceived as non-change (i.e. roles had existed also previously with different names). |
| 3 | Group: Process department, software projects | Feasibility of new tool | Evaluate new MT tool | Not achieved, action withdrawn | Requirements for tool acquisition procedure. |
| 4 | Group: Process department | Unawareness of current MT status, areas for improvement | Develop and distribute MT questionnaire | 57 responses gathered, current status clarified, no feedback back to projects, no action plan produced | Input provided for a larger test process improvement initiative launched in the new organization. |
| 5 | Organization: Management Internal group: Software project | New technology performance capability | Perform pilot project on new technology | Two working increments produced, no MT tools/methods evaluated | Process department was unable to deliver tools necessary for the pilot project: requirements for tool acquisition procedure. |
| 6 | Group: Process department, Individual: Researcher | Module testing practices will not be improved | Hire an outside researcher to evaluate reasons for the little progress made | 12 in-depth interviews performed, suspected reasons for the progress made reported to process department head and MT Core Team | Input provided for a larger test process improvement initiative launched in the new organization. |

### 5.2.3 Project closure: Changes achieved

While the target of the SPI initiative was to change the module testing practices of software projects this was not achieved within the 12-month period. However, whereas in case one the commitment net of the management contained a clear concern about usability issues, no indication of such a concern for module testing development is revealed by studying the open-ended interviews concerning the commitment net of the focal organization management (Table 11).

The only concern that indirectly relates to module testing development is the concern about the efficiency of new technology. SPI action #5 involved actions to match this concern. The management followed up the project closely. Weekly "news flashes" were delivered to a larger body of audience. The management communicated their expectations clearly. In addition to the performance evaluation, also the module testing tools, methods and processes should have been piloted. The pilot team managed to produce two working increments, thus achieving their most important goal. However, they failed to solve any module testing issues, even though an additional dedicated resource was included in the pilot at later stage to tackle specifically this problem, which, again, can be considered a sign of increased commitment. Yet, the pilot team is not to blame for this, as it was the process department that was not able to deliver the platform, tools and methods needed for testing within the time limits of the pilot project. This was also partially caused by problems with tool vendors (i.e., unacceptable licensing rules).

It was the commitment net of the process department that did show concern about the *development* of module testing practices. They were attempting to transfer this concern for software projects by defining new roles (#2). As stated earlier in the analysis, this transfer was not successful due to their perception that there was essentially nothing new in this role and that they did not want to take over the responsibility from the process department.

The module testing improvement effort lacked personnel dedicated to developing practices. The software projects had a clear view on the process department's role in this matter. It was up to their commitment net to solve these types of problems. Software projects wanted packaged support – including the tools, methods and processes needed for each separate environment. The process department was aware of this, but was not able to deliver support of the kind. They were understaffed due to an increased number of software projects and operating environments. Thus, the improvement strategies employed did not succeed in altering the software projects commitment net. Thus, no commitment to SPI was achieved apart from the process department. A software designer clarified this by stating:

> "Yes, but there is no such thing [module testing development] in the project [plan]. At least, as of yet it hasn't. It is stated [in the project plan] that the [cycle] time must be reduced. Probably something like that." [Software designer, interview]

Table 11. Case 2: Management's commitment net.

| Drivers | Commitment | | Outcomes | |
| --- | --- | --- | --- | --- |
| | Concerns | Actions | Intended | Unintended |
| Historical development<br> "Giving up old world, moving to new world"<br>Organization's state of will<br>Problems with self-developed tools<br>Technological advancements<br>Partnering, subcontracting and outsourcing pressures<br>Beliefs | Emerged technology, risks associated<br>Make sure projects stay on schedule<br>Produce a product that can be further developed<br>Use of commercially developed methods/tools<br>Concentration on core competence<br>Quality requirement: Down-time/year<br>To have right people in right projects<br>Internal free recruitment<br>Understanding the new technology (maintaining personal competence)<br>Maintaining trust and authority<br>Achieve 'high-enough' quality<br><br>Performance efficiency of new technology | Make management decisions<br>Strategy formation<br>Vision formation<br>Participate in product related steering<br>Group meetings<br>Hold personnel discussions<br>Follow up progress<br>Make resource decisions<br>Recruitment policy<br>Handle resistance<br>Own competence maintenance and development<br><br>Communicate expectations | Expectations clear: Test process automation | Module testing was prioritized lower [than other phases] |

Thus, software projects are constrained by the project plan. Their dominating concern is to meet the requirements stated in the plan. Other issues are treated as secondary. Table 12 summarizes the extent of change achieved in commitment nets at strategic and operational levels.

Table 12. Case 2: Changes achieved in actors' commitment nets.

| Commitment Net | Target actor | SPI outcome | Evidence |
|---|---|---|---|
| Strategic | Organization: Management | No concern about module testing was developed. In their opinion, it was up to the process department to develop module test processes. | Strategic commitment net does not contain concern with MT development. |
| Operational | Group: Software projects | No concern about module testing was developed. In their opinion, process department should be developing module test processes. | Reluctance to participate in MT development, new roles lost their significance in the new organizational structure. |

## 6. Results

The case analyses provide insight into how an SPI is driven through formation and reformation of commitment nets at different operating levels. The success of these SPI initiatives was differentiated in terms of how well the respective concern was transferred to the actors' commitment net. Fundamentally, SPI attempts to change the way software is produced. Thus, ultimately software projects (operational) and the commitment nets of individual designers (personal) should change. Seeing that SPI requires a strong support from the managerial levels, also changes in the strategic commitment net are called for. This section seeks to generalize the results of the study in terms of the contents of the personal, operational and strategic commitment nets.

6.1. Personal commitment nets

Each individual has a personal commitment net. It is argued that change efforts will be easier to perform if the contents of such a net are explicated. However, it is acknowledged that it is difficult to analyze the contents of an actor's commitment net, since personal commitment nets are likely to be personal also in their literary sense – they are not often willingly shared. A researcher is often faced with espoused theories rather than theories-in-use, i.e. the informant tells the researchers how matters should be rather than how they are. This was the case in the first case, in which the senior manager was explaining in detail how business objectives were to be met and how the organization would remain competitive, while the research team was trying its best to establish a relationship between business objectives and usability. In turned out that this particular manager had recently been attending lectures on these issues and was reflecting how these objectives should be met. In reality, the business plan had been formed several years ago together with a business consultant and had been long outdated. Data and methodology triangulation as well as field observations and participation in the activities of the case organization provides help in this regard. Also open interviews proved to be useful.
It was found that the role and the position held by the software practitioner in the organization are likely to determine the contents of the personal commitment net. Any changes to the role will bring about changes in the role related issues of the commitment net content. Individuals hold different beliefs about how software should be produced, how other actors should contribute to the work, etc. This is referred to as "perceived commitment net". Beliefs, in turn, influence the formation of concerns. Dybå [2000] has identified a number of key success factors for SPI. Among others, the concern about metrics was included. In case two, the metrics related beliefs were expressed by an experienced software designer as follows

> "[It is not about control], I think [module] testing is reasonable without tracking down faults. [...] I think that sometimes you get more [faults] and at other times less, some people get more [faults] than others. It depends on the type of feature that will be coded, how difficult it is and how many of them [difficult modules] there are. If we are implementing a simple feature, there will not be too many faults. It is all about that. The metrics do not make any difference. There will always be [faults]." [Software designer, interview]

The beliefs presented above are based on the observations and personal experience of the designer gained through having worked over six years in the industry. Introducing some – potentially time consuming – data collection activities to him would be difficult at first. But, if we wish to find out what his true concerns are, we could nourish these concerns and build upon this basis. He, among others, indicated that work practices should be efficient. Thus, if data collection activities are to become a target of his commitment, they have to contribute to the efficiency of his work at least to some degree.

The data collected has proven that past experiences have a strong influence on the concerns of the actors involved. Company A had undergone an organizational restructuring a year before the usability project was initiated. During the restructuring phase, a consultant had interviewed the employees about their work, and some of these employees were made redundant after these interviews. Since examining the current status of usability processes was considered a task of high priority, a number of employees were interviewed as a part of the assessment. In turned out that the main concern for most of employees was security, e.g. regarding their role and position in the organization. This was something the research team failed to realize early on. The security concern affected the assessment atmosphere and the information collected. A member of the management team suggested for the researchers to present the results in a positive light, e.g. involving proposals for consequent actions, which was done.

Regarding the concern creation for the process improvement initiatives studied, the data shows that task characteristics (challenge, importance and relation to actor's work tasks), on one hand, and concrete benefits gained, on the other hand, both affect the concerns and consequent actions at all actor levels. Any action in SPI, therefore, is likely to affect the operational and personal levels – and ultimately also the strategic level. It is, however, very difficult to observe or to analyze the effect on the strategic level in terms of return-on-investment. Only few studies have been conducted trying to explicate how much benefit an organization has gained of any technological innovation, including process innovations such as the CMM [Glass 1999]. The research effort in case 2 was designed to discover problems and to propose solutions in module testing development endeavors. All the key personnel were interviewed in this context, but the results were not reported beyond those immediately concerned. A more profound impact would have been gained if the results had been reported also to the designers, project managers and managers involved.

If SPI did not prove its position as a meaningful commitment target, what did? The interview data shows that at the personal level, the concerns shared widely were current work task, meeting deadlines, career and competence development, quality of one's work and work efficiency. These findings indicate that any new innovation made in software engineering should nurture these concerns. Software practitioners interviewed did not want to "waste their time". Thus, tools that are overly difficult to use, methodologies that are difficult to grasp and processes that are too detailed present themselves an attack against the natural concerns of the software engineer. These findings are in line with Glass [1999], who has studied the payoffs of software technologies. He has reported that although the cleanroom error-removal technique is extremely efficient, people dislike the approach. Despite being important and meaningful, the used approaches and tools should also bring out a sense of fulfillment and satisfaction enhancing the creativity in one's job [McGuire 1996]. Humphrey's [1995] approach to personal capability development – the personal software process (PSP$^{SM}$) methodology – defines what the concerns of every software practitioner should be and how to implement these. Thus, PSP can be seen as an ideal type of personal commitment net, involving concerns about the process and product quality. Table 13 presents the findings concerning the personal commitment net level, worked out using the case data.

Table 13. Personal commitment net.

| Drivers | | Commitment | |
|---|---|---|---|
| External | Internal | Concerns | Actions |
| Operating context (software engineering profession) <br> Organization's state of will <br> Sub cultures <br> Marketplace | Role <br> Position <br> Beliefs <br> Past experiences <br> Expectations <br> Task characteristics <br>   Challenge <br>   Importance <br>   Fulfillment (satisfaction) | Current work task <br> Meeting deadlines <br> Career development <br> Quality of work <br> Work efficiency (also productivity) <br> Competence development <br> Security <br> Organization's well-being | Role/position related <br> Role/position unrelated |

6.2. Operational commitment net

The cases studied evidenced that the core target of improvement is, by necessity, the operational commitment net in the organization, i.e. the software project itself. In the first case, the SPI approach had to be changed in order to better meet the needs of the software project. In the second case, the project manager was the leader of the module testing core team. However, having the project manager involved is no guarantee that the operational level commitment net will eventually change. The same software project manager felt that it was his responsibility to "protect" the project members from outside turbulence, referring to SPI activities. Our findings here are in line with the SPI literature. Nielsen and Nørbjerg [2001] argue that software project managers are facing contradictory demands and conflicting interests, uncertainty and change, which they have to be able to cope with during the development project. Nielsen and Nørbjerg interpret project managers' actions as an efficient means to steer the project development through this unstable environment, and not as a sign of low maturity, as would be understood through a software capability assessment model lacking the operating context. Software process specialists are facing contradictory demands as well. In case two, the process people were expected to provide "packaged" support for the projects.

> "Of course, that is how it should be – a [support] package should be offered, but when you do these things alone, there are simply no opportunities for such. There are already dozens of tools available due to different [operating] environments and others." [SPI project manager, interview]

Thus, several items that appear in the operational commitment net were identified from the case material. The external drivers operating at project level are the state of will of the organization, specific sub cultures and the marketplace. These drivers are something that a project does not have direct influence on. The internal drivers include past experiences, shared beliefs and the project manager, who plays a strong role in the cases studied. Shared beliefs have emerged over time – however not necessarily reflecting the reality. The commitment at the operational level is toward to the project itself. Issues that make the project succeed or fail are high on the list of main concerns. The concern about process compliance in terms of procedure compliance was also identifiable: In case two, the project manager explained how the inspection process is a routine activity for them, even though the inspection reports are rarely used, if ever. Inspection activity thus appears as the standard procedure of the company for accepting a piece of work for distribution or evaluation. In fact, it was stated by the project manager that a piece of work that had not been inspected was not considered valid. Thus, the compliance for the process exists, but not in a way as it is described in quality manuals. Therefore, the compliance does not concern the process but the procedure itself. No concern about the capability for SW development was directly identifiable or observable at the project level. Instead, the quality of the outcome of their work was. Projects seem to perceive SPI as something that is unrelated to their role in the organization. To them, SPI falls within the scope of activity of the quality department. Their own concern should be the welfare of the software project.

Table 14. Operational commitment net.

| Drivers | | Commitment | |
| --- | --- | --- | --- |
| External | Internal | Concerns | Actions |
| Operating context | Past experiences | Project | Project related |
| Organization's state of will | Shared beliefs | Technology | Project unrelated (e.g. SPI) |
| Sub cultures | Project manager | Project deadlines | |
| Marketplace | | Quality of outcome | |
| | | Procedure compliance | |

6.3. Strategic commitment net

Strategic commitment nets have a powerful impact on the contents of lower level nets. Case data strengthens the evidence that SPIs are rarely performed in vacuum. When process innovations are studied, a larger picture should be developed. In the second case, several strategic decisions had been made: the use of commercial tools, i.e. no more in-house tool development, reduce cycle time and outsource part of old product development. These actions come from concerns like the need to stay competitive, secure production efficiency, round-the-clock development, etc. The need of reducing cycle time led to the adoption of the incremental software development paradigm, which in turn caused major changes in all operational and personal commitment nets. Also the emerging new technologies caused the need to change the implementation technologies as well. In both organizations, the emerging new product appeared to tie a great deal of organization resources. In the first case, the whole usability project was centered on ensuring the success of the new product in terms of usability. Thus, while lower level process improvement may occur, it is always jeopardized by strategic concerns. For this reason, SPI authors are requiring that all SPI activities should be made a business case [cf. Jones 1999].

Both cases strengthen the evidence that business decisions and changes in the organization can abruptly seize any improvement efforts, thus emphasizing the sensitivity of SPI activities to external environment. Nielsen and Nørbjerg [2001] found that there is a constant change in a project environment where the organizational structure is changing as new managers are hired and others leave. They argued that the fluctuations in product sales or in the size of market segments might reshape the whole company – leading to changes in the commitment nets at all levels. Table 15 presents the contents of the strategic commitment net.

Table 15. Strategic commitment net.

| Drivers | | Commitment | |
|---|---|---|---|
| External | Internal | Concerns | Actions |
| Operating context<br>Marketplace<br>Economy<br>Technological development | Organization's culture<br>History<br>Company demographics | Competitiveness<br>Development capability<br>Company values<br>Product quality<br>Company image<br>Organizational efficiency<br>Market segment<br>Product sales<br>Shareholder satisfaction | Strategic decisions |

## 7. Discussion

This research has shown how commitment can be found and explicated in software organizations. Our study sheds light on how the commitment to a process improvement project is generated and what the obstacles are along the way. The commitment nets model used did not presuppose that commitment to SPI would necessarily come about. It was evident that it was other commitments that dominated the daily work of software practitioners. In the results section, these commitment targets were identified at the different organizational levels in generic terms.

In the light of these two cases it can be argued that SPI in itself is not a natural commitment target, in the way current tasks and accepted responsibilities are. In our cases, software designers mainly concerned themselves about their task contents, work efficiency, keeping up with the schedule and the quality of their work. Therefore, SPI should be integrated into the daily work of software practitioners. Otherwise, it will be considered part of some other commitment nets. Our finding here is in line with other researchers [Hall and Wilson 1997; Sharp, et al. 2000] arguing that the software engineering practice community is not lacking in commitment to quality. The software practitioners in our cases showed a lot of interest in issues that would benefit their work and make it more efficient.

However, our case data reveals that attempts are still often made to bring quality from outside. In the second case, a separate quality audit was performed. One of the major findings of this audit was that a tool that was specified by the management to be used in module testing was not used, because it was quite simply unsuitable for the task. However, it the tool was found suitable for the integration test phase, which was why is was thought also to benefit the module testers. The module testers, for their part, made "a silent agreement" not to use the tool and resorted to their old practices.

How then should practitioners incorporate new technologies into their software engineering processes? It is suggested that although organizational mandates are often used [Chau 1996] they tend to fall short regarding their impact on organizational effectiveness. Strategic decisions do direct the organizational commitment nets at all levels. Our cases show that these directions are often understood, even though they are not always willingly accepted. In the second case, the decision to outsource part of the development was made. The outsourcing time was kept to minimum, which caused dissatisfaction at lower levels. However, professionals seemed to accept that "the Old World" has to be given up when "the New World" arrives. Thus, in spite of the dissatisfaction all the effort was directed to make it happen.

Our case data also indicates that grand scale improvement initiatives, which may begin with a one-week assessment, may belong to history now. SPI is clearly in its transition phase. Emerging process technologies under the umbrella of agile software development [Cockburn 2002] have drawn a lot of attention and impressive results have been achieved. Agile solutions such as extreme programming [Beck 1999] focus on individuals and small teams and their capabilities. This solves some of the problems that were observed in the two SPI cases, such as the one concerning the role of process department in a software organization. Agile technologies also seek to embed the process improvement activities into the daily work of software professionals as a natural part of the work routines, as suggested by Yamamura [1999]. The attempt to make continuous SPI as natural as possible sets new directions for tool vendors also. Emerging tools should incorporate the possibility to improve the process on the basis of experience and collected data. In the two cases of ours, both organizations experienced difficulties in finding competent staff for the process department. According to a manager interviewed they would traditionally place a junior software engineer in the process department to learn about the organization. After some time this person would be moved to a software project. Thus, it is no surprise that collaboration tends to be difficult.

## 8. Conclusions

While organizations continue making considerable investments in SPI efforts, it has been suggested that nearly two thirds of these efforts are likely to fail [Debou 1999]. Commitment has been argued to play a major role in determining the level of success achieved in SPI. Still, studies addressing the complexities of commitment have been scarce.

This paper has made an attempt to fill this gap by developing a research model for studying the existence, development and interplay of commitment in SPI. This paper has suggested that software organizations operate through what can be called strategic, operational and personal commitment nets. These nets consist of actors, drivers, concerns, actions, commitment, and outcomes. The commitment net research model is currently being used as an analytical tool in studying the development of commitment in industrial SPI cases. It has been shown by this research that SPI is driven through formation and reformation of commitment nets. As a result, the contents of personal, operational and strategic commitment nets have been laid out as they have been extracted from the case data. The results indicate that when a single SPI initiative is studied, a broader view on the entire organization is needed, since SPIs are rarely carried out in vacuum. This study also sheds light on the role of SPI in the future, especially concerning its validity and usefulness.

This paper provides researchers and practitioners alike with new insights into commitment related phenomena. The ideas presented in this paper are grounded on empirical data. The proposed model serves as an operational vehicle for bringing out commitments, their relations and provides a typology based on common aspects of commitment for dealing with process related issues.

### Acknowledgments

The authors would like to thank prof. Veikko Seppänen for his contribution to the research model, Ms. Netta Iivari for the active involvement in the analysis of the first case, Mr. Timo Kucza for his helpful ideas and three anonymous reviewers for their comments on the early version of the paper. The authors are grateful to the participating companies for their time, assistance and feedback. This work has been financially supported by Infotech Oulu Graduate School.